\documentclass[journal,hidelinks]{IEEEtran}

\usepackage[T1]{fontenc} 
\usepackage{amsmath}
\usepackage[cmintegrals]{newtxmath}
\usepackage{bm} 
\usepackage{amsfonts} 
\usepackage{graphicx} 
\usepackage{hyperref}
\usepackage{color}
\usepackage{comment}
\usepackage{dcolumn}
\usepackage{bm}
\usepackage{microtype}
\usepackage{subfig}
\usepackage[noadjust]{cite}

\newcommand{\braket}[2]{\ensuremath{\left\langle #1 | #2 \right\rangle}}

\begin{document}

\title{The Resilience of Hermite- and Laguerre-Gaussian Modes in Turbulence}
%
%
%

\author{Mitchell~A.~Cox,~\IEEEmembership{Member,~OSA,}
        Luthando~Maqondo,
        Ravin~Kara,
        Giovanni~Milione,\\
        Ling~Cheng,~\IEEEmembership{Senior Member,~IEEE,}
        and~Andrew~Forbes,~\IEEEmembership{Fellow,~OSA}
\thanks{M. Cox and L. Cheng are with the School of Electrical and Information Engineering, L. Maqondo, R. Kara and A. Forbes are with the School of Physics, both at the University of the Witwatersrand, Johannesburg, South Africa. G. Milione is with the Optical Networking and Sensing Department, NEC Laboratories America, Inc., Princeton, New Jersey 08540, USA.  mitchell.cox@wits.ac.za.}
\thanks{Manuscript received ...; revised ...}}

\markboth{Journal of Lightwave Technology,~Vol.~?, No.~?, ?}%
{Shell \MakeLowercase{\textit{et al.}}: Bare Demo of IEEEtran.cls for IEEE Journals}


\maketitle

\begin{abstract}
Vast geographical distances in Africa are a leading cause for the so-called ``digital divide'' due to the high cost of installing fibre. Free-Space Optical (FSO) communications offer a convenient and higher bandwidth alternative to point-to-point radio microwave links, with the possibility of re-purposing existing infrastructure. Unfortunately, the range of high bandwidth FSO remains limited. While there has been extensive research into an optimal mode set for FSO to achieve maximum data throughput by mode division multiplexing, there has been relatively little work investigating optical modes to improve the resilience of FSO links. Here we experimentally show that a carefully chosen subset of Hermite-Gaussian modes is more resilient to atmospheric turbulence than similar Laguerre-Gauss beams, theoretically resulting in a 167\% theoretical increase of propagation distance at a mode dependent loss of 50\%.
\end{abstract}


%
\IEEEpeerreviewmaketitle

\section{Introduction}
\label{sec:intro}

Communication in the modern era was revolutionised by the invention of the electric telegraph and subsequently the conventional telephone in 1837 and 1876 respectively. The use of the copper and microwave radio infrastructure built for telephones eventually led to the Internet, which is now ubiquitous. In the 1970s, fibre optic communications technology began to replace the existing copper back haul infrastructure due to the fact that it is cheaper, higher bandwidth and longer range \cite{winzer2014a}. In terms of the access network it is clear that consumers were delighted with the service provided by copper wires and even today many Internet users still make use of ADSL instead of Fibre To The Home (FTTH), which is often unavailable. The cost and complexity of upgrading the historical copper infrastructure has been a limiting factor in the deployment of FTTH. For instance, in 2017 according the the FTTH/B Global Rankings from the IDATE for FTTH Council in Europe, the household penetration of FTTH was approximately 17\% in the Netherlands, 14\% in the United States and only 4\% in South Africa. All other continental African countries have less than 1\% FTTH penetration.

In developing countries such as those in Africa, access to the Internet with bandwidth sufficient for audio or video streaming, for example, is rare. Africa has 16\% of the world's population but only 4\% of the internet users. This is the basis of the so-called ``digital divide'' and is due to various factors both socio-economic and geographic \cite{lavery2018}. Using Africa as an example, the distances between human settlements and existing fibre infrastructure can be immense and given that the installation of fibre is expensive, bridging this digital divide in Africa is a significant challenge.

Existing internet infrastructure in Africa, where fibre is not available, is provided by satellite and terrestrial microwave links. Terrestrial microwave links consist of point-to-point, line of sight towers typically situated tens of kilometres apart, depending on the terrain. For instance, on flat ground, two 50~m high towers could be separated by approximately 40~km before the curvature of the Earth becomes an issue. Carrier-grade microwave technology is able to sustain throughputs in the region of 10~Gbps per link. Satellite connectivity is also limited by high latencies, low bandwidth and high cost. 

Free Space Optical (FSO) communication technology may be a viable technology to provide high bandwidth without the expense of installing long distance fibre back haul \cite{lavery2018,Zhao2015}. Existing high-sites and towers that are currently used for microwave links could be retro-fitted with FSO which can operate in parallel to the existing infrastructure to maintain high availability albeit at lower capacity. This may be seen as an interim technology to bridge the gap until higher bandwidth fibre is installed, in the case of large town and cities. Unfortunately, existing commercial FSO solutions have a limited range of about 2~km at several Gbps. There have been several demonstrations of extremely high bandwidth FSO communication links as well as numerous theoretical investigations using Mode Division Multiplexing (MDM) \cite{Willner2017,Zhao2015,Trichili2016,Chen2016a}. Unfortunately, the long distance challenge for MDM-FSO has not been solved, although there have indeed been promising demonstrations \cite{Lavery2017,Krenn2014,Krenn2016,zhao2016linkdemo,Ren2016}.

One of the primary issues with FSO is turbulence which results in what is typically called optical scintillation and manifests as the random fluctuations (or fading) of the received signal's intensity. For MDM, turbulence also distorts the wavefront resulting in crosstalk, thus reducing the link capacity. It is difficult to mitigate the effects of turbulence and common strategies are to use signal processing techniques such as MIMO and strong forward error correction as well as adaptive optics \cite{Willner2017}. Channel diversity is another well-known technique where multiple transmitters or receivers are spatially separated to reduce the probability of errors \cite{Navidpour2007a}. It has been demonstrated that modal diversity, where the spatial separation is achieved using modes instead of physical separation is viable \cite{cox2018,Huang2018,mehrpoor2015}. 

The question of whether there are optical modes that are more resilient to atmospheric turbulence than a standard Gaussian mode is pertinent as this would be an effective way to passively increase the range of a FSO link. Typical beams that have been well investigated in turbulence are Orbital Angular Momentum (OAM) or Laguerre-Gauss (LG) modes, Bessel-Gauss modes and also vector modes with spatially varying polarisation, with varying degrees of success \cite{Cheng2009a,Aksenov2016a,Mphuthi2018bessel,Cox2016,Gu2009}.  In general it has been found that higher order modes are indeed more resilient than Gaussian modes \cite{Aksenov2016a}. In the presence of a restricted aperture, LG (or rather OAM modes) modes have a larger information capacity than HG modes \cite{Restuccia2016}, but if this is not the case then HG modes are a promising candidate as they are robust against tip/tilt aberrations \cite{Xie2015,Pang2018HGLG,ndagano2017}.

Given this observation, assuming no aperture restrictions, are HG modes (or a subset of HG modes) more resilient than LG modes in atmospheric turbulence and what gains in terms of propagation distance can be expected? We hypothesize that since the dominant effect of turbulence is tip/tilt \cite{Noll1976}, HG modes will indeed be more resilient than LG modes, however, it is unknown whether the higher order effect of turbulence will counteract this gain.

In Sec.~\ref{sec:background} we provide a summarised background of HG and LG modes, as well as a convenient, phase-only approximation of an HG mode which we call a Binary HG (BHG) mode. Orthogonality is critical to MDM and so we briefly discuss and show how the BHG modes are not always orthogonal as one might naively assume. Furthermore, we provide a short overview of atmospheric turbulence and how it affects the orthogonality of these mode sets. Due to their different phase structure, it is logical to assume that different modes will be affected by atmospheric turbulence in different ways. It would be highly advantageous to a MDM system if a certain set of modes exhibited lower losses or less crosstalk. An experimental setup and methodology to determine whether (B)HG modes are ``better'' than LG modes in these respects is described in Sec.~\ref{sec:experiment} with results and discussion in Sec.~\ref{sec:results}. The paper is concluded in Sec.~\ref{sec:conc}.

\section{Preliminaries}
\label{sec:background}

For the reader's convenience, we provide a brief background of HG, Binary HG and LG modes, their orthogonality and how the impact of atmospheric turbulence on the modes is typically measured. 

\subsection{Laguerre- and Hermite-Gaussian Modes}

\begin{figure}[t]
  \centering
      \includegraphics[width=1.0\linewidth]{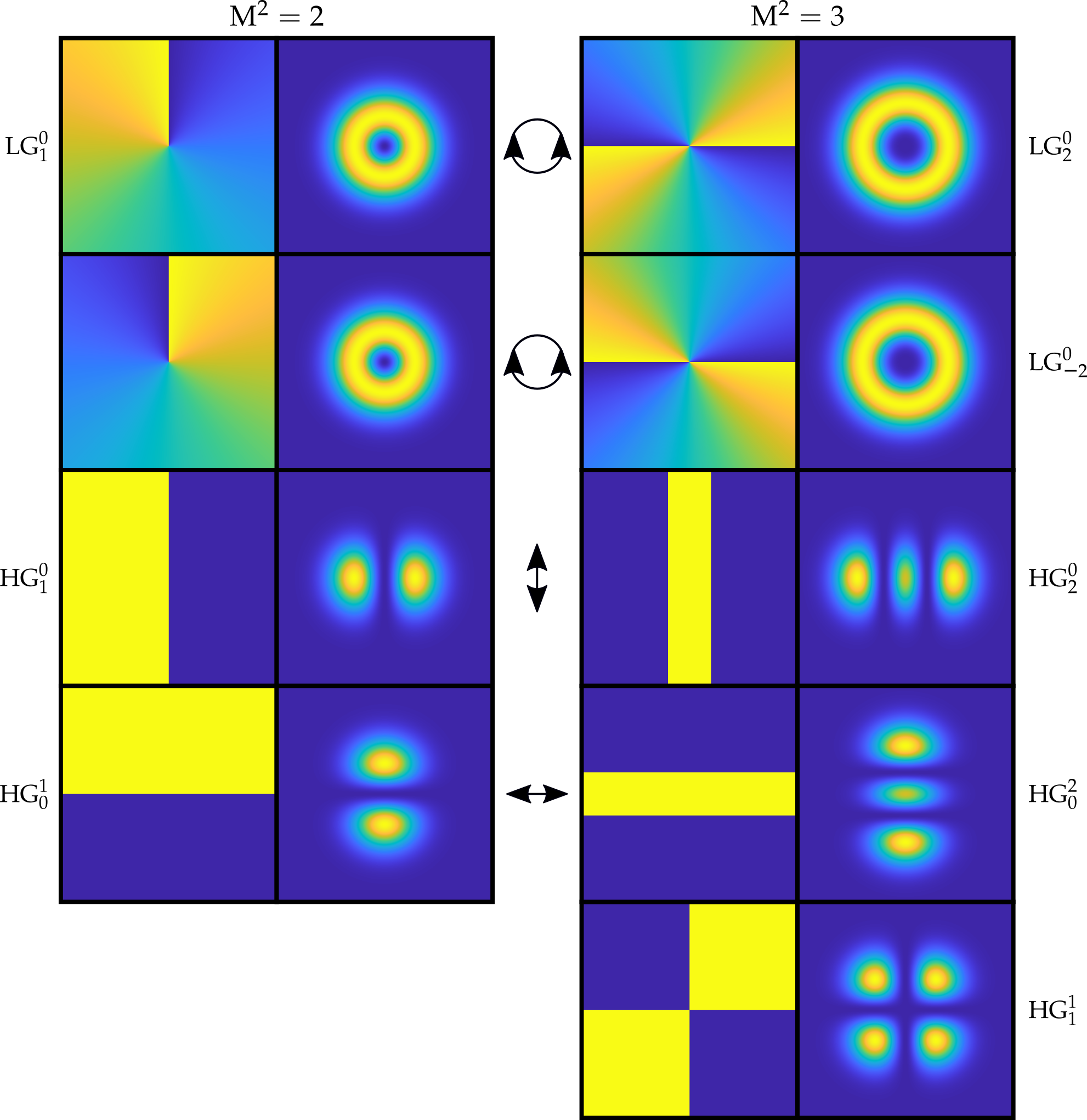}
  \caption{\label{fig:modes}Modes used in the experiment, with wavefront to the left and intensity to the right of each, corresponding to Tab.~\ref{tab:modes}. The arrows in the centre indicate the translational (or rotational) axis where the mode is robust and will still be detected correctly. The HG$^1_1$ mode does not have a robust axis.  }
\end{figure}

The higher order solution to the paraxial wave equation in cylindrical coordinates is called a Laguerre-Gaussian (LG) mode and in Cartesian coordinates is called a Hermite-Gaussian (HG) mode, with examples of the mode intensity and phase in Fig.~\ref{fig:modes}. The modes within each of these sets are orthogonal to each other, making them useful for multiplexing. When the beams are generated with a Spatial Light Modulator (SLM) we encode the modes at $z=0$, resulting in Eqs.~\ref{eq:LG} and \ref{eq:HG} respectively.
\begin{equation}
\label{eq:LG}
\begin{aligned}
&U^{\mathrm{LG}}_{\ell,p}(r,\phi) = \\&C^{\mathrm{LG}}_{\ell,p} 
\left(\frac{r\sqrt{2}}{w_0} \right)^{\!\!|\ell|} 
L_p^{|\ell|}\!\!\left(\frac{2r^2}{w^2_0}\right)
\exp\!\!\left(-\frac{r^2}{w^2_0} \right)
\exp (-i\ell\phi)
\end{aligned}
\end{equation}
\noindent where $\ell$ and $p$ are the mode indices, $C^{LG}_{\ell p}$ is a normalisation constant and $L_p^{|\ell|} (\cdot)$ is the generalised Laguerre polynomial and
\begin{equation}
\label{eq:HG}
\begin{aligned}
&U^{\mathrm{HG}}_{n,m}(x,y) = \\&C^{\mathrm{HG}}_{n,m} \, H_n\!\!\left(\frac{\sqrt{2} \,x}{w_0}\right) H_m\!\!\left(\frac{\sqrt{2} \, y}{w_0}\right) \exp\!\!\left(-\frac{x^2+y^2}{w_0^2} \right) 
\end{aligned}
\end{equation}
\noindent where $n$ and $m$ are the mode indices, $C^{\mathrm{HG}}_{nm}$ is a normalisation constant and $H_n (\cdot)$ and $H_m (\cdot)$ are Hermite polynomials of order $n$ and $m$ respectively. 

We can encode HG modes without amplitude (i.e. only phase) information, resulting in what we call a Binary HG (BHG) mode. A BHG mode has only two phase values (similar to an HG mode) of $0$ and $\pi$ with the amplitudes constrained to be either $0$ or $1$. This is achieved simply using Eq.~\ref{eq:BHG}:

\begin{equation}
\label{eq:BHG}
U^{\mathrm{BHG}}_{n,m}(x,y) = \frac{1}{2} + \frac{1}{2}\text{sign} \left[ U^{\mathrm{HG}}_{n,m}(x,y) \right] 
\end{equation}

We consider BHG modes in addition to standard HG modes because they can easily be created and decomposed with a refractive element. Unfortunately, while LG and HG modes form an orthonormal basis, the BHG modes are not always orthogonal:
\begin{equation}
\label{eq:BHGOrthogonality}
\begin{aligned} 
\braket{U^{\mathrm{BHG}}_{n,m}}{U^{\mathrm{BHG}}_{n',m'}} = \left\{
        \begin{array}{ll}
            0 & m \neq m' \ \text{and $m$ is even}  \\
            0 & n \neq n' \ \text{and $n$ is even}  \\
   			c & \text{elsewhere},  
        \end{array}
   \right.
\end{aligned}
\end{equation}
where $c$ is some normalisation constant. The HG modes are also not always orthogonal to BHG modes, as an example, the orthogonality result is an asymmetric matrix when you keep one index equal to zero, $m=0$ or $n=0$. 

Since the detection system for higher order modes often makes use of a hologram which works as a matched filter or inner product measurement, the alignment of the incoming beam onto the hologram is critical. Adaptive optics are effective at reducing mode dependent loss (MDL) and mode-crosstalk \cite{Li2018b}. This is because adaptive optics, and indeed even a simple tip/tilt mirror, are able to correct any misalignments in addition to wavefront corrections. 

Figure~\ref{fig:modes} shows the phase and intensity of several different LG and HG modes. For an LG mode to be correctly detected it must be aligned so that the vortex of the mode is centred on the corresponding phase singularity encoded on the hologram. Any translation (called tip or tilt) from this overlap will manifest as crosstalk into neighbouring modes, however, because the beam is radially symmetric, rotation of the beam will not affect the detection. HG modes are symmetric with respect to a horizontal and/or vertical axis. Unlike LG modes, rotation of an HG mode will adversely affect the integrity of the measurement, however, it is clear from the figure that HG modes which are symmetric about only one axis will be resilient to translation along that axis. For instance, the HG$^1_0$ mode can move left or right without any detection error whereas the HG$^1_1$ mode must be well aligned similar to the LG modes. 

The completeness property of the LG and HG bases allow us to express any element of one basis as a linear combination of elements from the other basis using the transformation relations in Eqs.~\ref{eq:LGtransform} and \ref{eq:LGtransformB} \cite{ONeil2000}. Note that here the LG modes have been written in terms of $n$ and $m$, which are indices typically used for HG modes. The usual indices can be recovered as $\ell=n-m$ and $p=\mbox{min}(n,m)$. 
\begin{equation}
\label{eq:LGtransform}
U^\mathrm{LG}_{n,m}(x,y,z)=\sum_{k=0}^N i^kb(n,m,k)U^{\mathrm{HG}}_{N-k,k}(x,y,z)
\end{equation}
\begin{equation}
\label{eq:LGtransformB}
b(n,m,k)=\left[\frac{(N-k)!k!}{2^Nn!m!}\right]^{\!1/2}\frac{1}{k}\frac{d^k}{dt^k}[(1-t)^n(1+t)^m]|_{t=0}
\end{equation}
where $N=n+m=2p+|\ell|$ is the order of the beam. It has been shown that because of this unitary transformation between all LG and HG modes, there is no capacity benefit in atmospheric turbulence when the average of the entire basis is considered \cite{Chandrasekaran2014}. This appears contradictory to the hypothesis presented in this work, however, we propose that a carefully chosen subset of HG modes is not subject to the work in \cite{Chandrasekaran2014}. In addition, there is a diversity benefit due to differences in the wavefront of the modes and short-term changes in atmospheric turbulence \cite{cox2018}. 

\subsection{Atmospheric Turbulence}
\label{subsec:turb}
When a laser beam propagates through the atmosphere it encounters spatially and temporally varying refractive indices, mainly due to random temperature variations and convective processes. This randomly aberrates the beams wavefront. The Kolmogorov model for turbulent flow is the basis for many contemporary theories of turbulence and is able to relate these temperature fluctuations to refractive index fluctuations \cite{kolmogorov1941}. In the Kolmogorov model, the average size of the turbulent cells are specified by a so-called inner scale, $l_0$, which is typically on the order of millimetres and an outer scale, $L_0$, which is on the order of meters \cite{Andrews2005}. Kolmogorov turbulence assumes $l_0 = 0$ and $L_0 = \infty$, thus the model's simplicity.


While models of turbulence typically only provide statistical averages for the random variations of the atmosphere, in most cases this is sufficient. The power spectral density of the refractive index fluctuations given by the Kolmogorov model is described by
\begin{equation}
\phi_n(\kappa) = 0.033 C_{\!n}^2 \kappa ^{-11\!/3} \quad \mathrm{for} \quad  1/L_0 \!\ll\! \kappa \!\ll\! 1/l_0,
\end{equation}
\noindent where $\kappa = 2\pi(f_x \cdot \hat{x} + f_y \cdot \hat{y})$ is the angular spatial frequency and $C_{\!n}^2$ is the refractive index structure parameter. We can use this to generate individual snapshots of turbulence in the form of phase screens with appropriate statistics \cite{HamadouIbrahim2013}. Instead of $C_{\!n}^2$, turbulence strength is often specified using the Fried parameter \cite{Fried1965}, commonly known as the atmospheric coherence length,
\begin{equation}
\label{eq:r0fromCn2}
r_0 = 0.185 \left(\frac{\lambda^2}{C_{\!n}^2 L}  \right)^{\!\!-3/5}
\end{equation}
where $\lambda$ is the wavelength and $L$ is the propagation distance. Furthermore, a more general parameter to specify turbulence strength is the Strehl Ratio (SR), which is the ratio of the average on-axis beam intensity with, $I$, and without, $I_0$, turbulence and is given by
\begin{equation}
\label{eq:sr}
\mathrm{SR} = \frac{\langle I(0,L) \rangle}{I_0(0,L)} \cong \frac{1}{[1+(D/r_0)^{5/3}]^{6/5}},
\end{equation}
where $D$ is the aperture diameter. In summary, turbulence leads to scintillation, beam wandering and other effects, and is the reason why the on-axis beam intensity, $I$, is reduced on average. The intensity reduction of the individual modes is called MDL where instead of the overall intensity of the beam, $I$, we use the intensity of individual modes, $S_i$,
\begin{equation}
\label{eq:mdl}
\mathrm{MDL}_i = 1 - \frac{S_i}{S_0},
\end{equation}
where $S_0$ is the intensity of mode $i$ in the absence of turbulence. Typically, the energy in the individual modes is spread to neighbouring modes, as mentioned in the introduction. This mode crosstalk with respect to mode $i$ is defined as the fraction of the total intensity not in mode $i$:
\begin{equation}
\label{eq:crosstalk}
C_i = 1 - \frac{S_i}{\sum_j S_j},
\end{equation}
where $\sum_j S_j$ is the sum of the intensities in all the modes, including mode $i$.

\section{Experimental Setup and Methodology}
\label{sec:experiment}

\begin{figure}[t]
  \centering
      \includegraphics[width=1.0\linewidth]{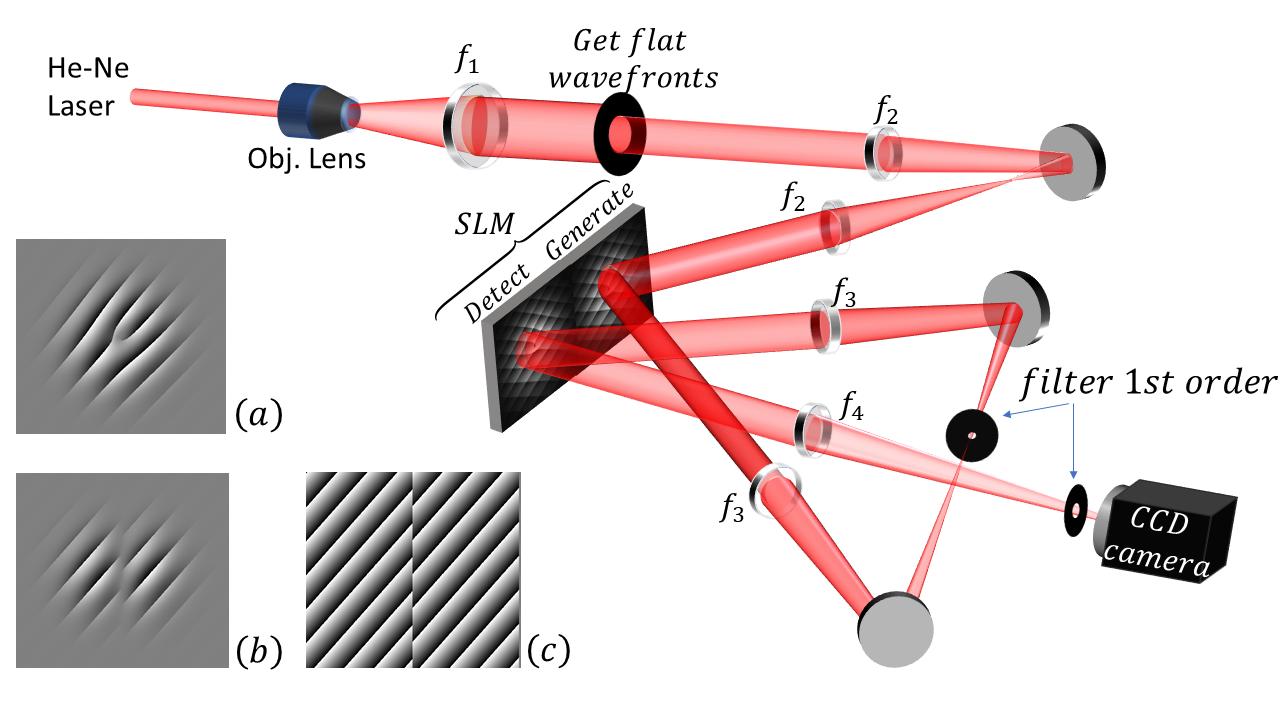}
  \caption{\label{fig:setup}Modal decomposition setup with sample insets of the holograms used: (a) LG$_{1}^{0}$, (b) HG$_{1}^{0}$ and (c) BHG$_{1}^{0}$. }
\end{figure}

In order to determine whether the MDL and crosstalk characteristics of HG, BHG and LG modes is different and ultimately if one mode set performs better than the others in Kolmogorov turbulence, we use a modal decomposition setup which makes use of a Spatial Light Modulator (SLM) \cite{flamm2009}. The SLM is used to create the required mode, aberrate that mode using emulated Kolmogorov turbulence, and finally perform a modal decomposition. The MDL and crosstalk for each set is then calculated.

A diagram of the experimental setup is shown in Fig.~\ref{fig:setup}. A 633~nm laser beam from a Helium-Neon laser is expanded using an objective lens and lens $f_{1}$. The flat wavefront from the small central region of the expanded beam is selected using an aperture and is imaged onto the SLM screen using a $4f$ system. A HoloEye Pluto SLM is divided into two halves, for two separate holograms. The first hologram is used to modulate the incoming flat beam into the desired mode as well as to add turbulence. The resulting field is imaged to the second half of the SLM where modal decomposition is performed. A camera is placed at the focal point of lens $f_{4}$ to measure the on-axis intensity, which represents $S_j$ in Sec.~\ref{subsec:turb}, Eqs.~\ref{eq:mdl} and \ref{eq:crosstalk}. 

For each mode in each set, HG, BHG and LG, modal decomposition was performed for one hundred random turbulence screens for each Strehl Ratio from 1.0 (no turbulence) to 0.1 (strong turbulence). The results were then averaged according to Strehl Ratio. To ensure a fair comparison, we used modes with the same beam propagation factor, $M^2 = n + m + 1 = 2p + |\ell| + 1$, for (B)HG and LG modes respectively. Table~\ref{tab:modes}, below, details which modes are used for each set, graphically shown in Fig.~\ref{fig:modes}. Note that for $M^2 = 3$, we also include a case which does not include the ``symmetrical'' (B)HG$^1_1$ mode, denoted by the asterisk. This case is considered because the HG$_1^1$ mode does not have the same benefit of tip/tilt invariance as the other (B)HG modes in the set. It is expected that when this mode is included, the average performance of the set will be more similar to the performance of the LG modes.

\begin{table}[th]
\centering
\caption{\label{tab:modes} Mode sets used in the experiment.}
\begin{tabular}{ccc}
$M^2$ & LG $(\ell,p)$ Modes & (B)HG $(n,m)$ Modes \\
\hline
$2$     & (-1,0), (1,0) & (1,0), (0,1)    \\
$3$     & (-2,0), (2,0) & (2,0), (1,1), (0,2)  \\
$3^*$   & (-2,0), (2,0) & (2,0), (0,2)        
\end{tabular}
\end{table}

\begin{figure}[th]
  \centering
      \includegraphics[width=1.0\linewidth]{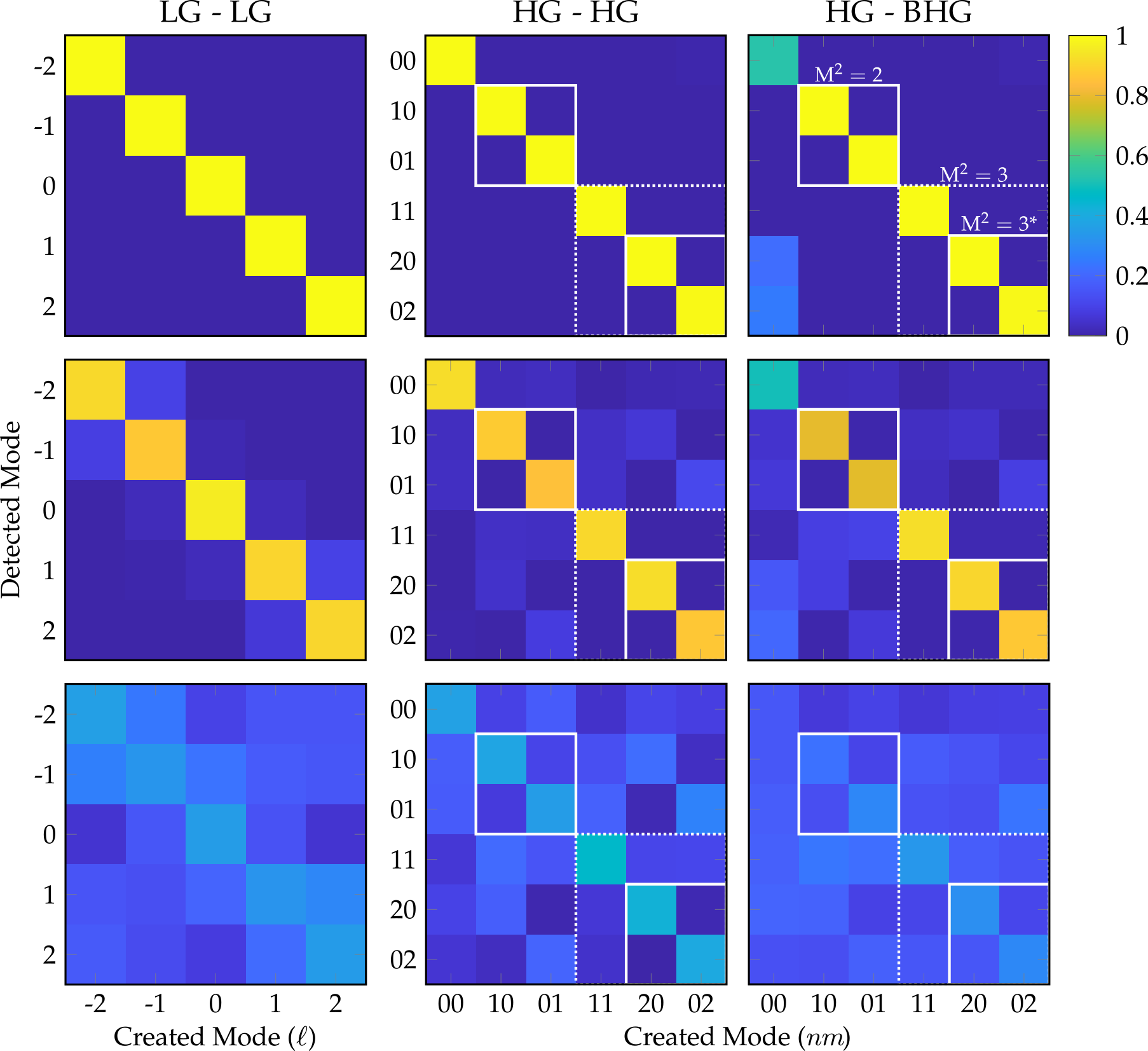}
  \caption{\label{fig:verification}Experimental setup verification showing the crosstalk matrices for the various mode sets at different turbulence strengths. Top: no turbulence (SR=1.0). Middle: mild turbulence (SR=0.7). Bottom: strong turbulence (SR=0.1). Mode groups corresponding to modes in Fig.~\ref{fig:modes} and Tab.~\ref{tab:modes} are highlighted as it is clear that there is relatively little crosstalk within each group.  }
\end{figure}

The experimental setup was verified by performing a modal decomposition for each mode set without turbulence. As shown in Fig.~\ref{fig:verification}, there is no crosstalk except where it is expected in the case of the BHG modes.

\section{Results and Discussion}
\label{sec:results}

\begin{figure*}[th]
  \centering
      \includegraphics[width=0.9\linewidth]{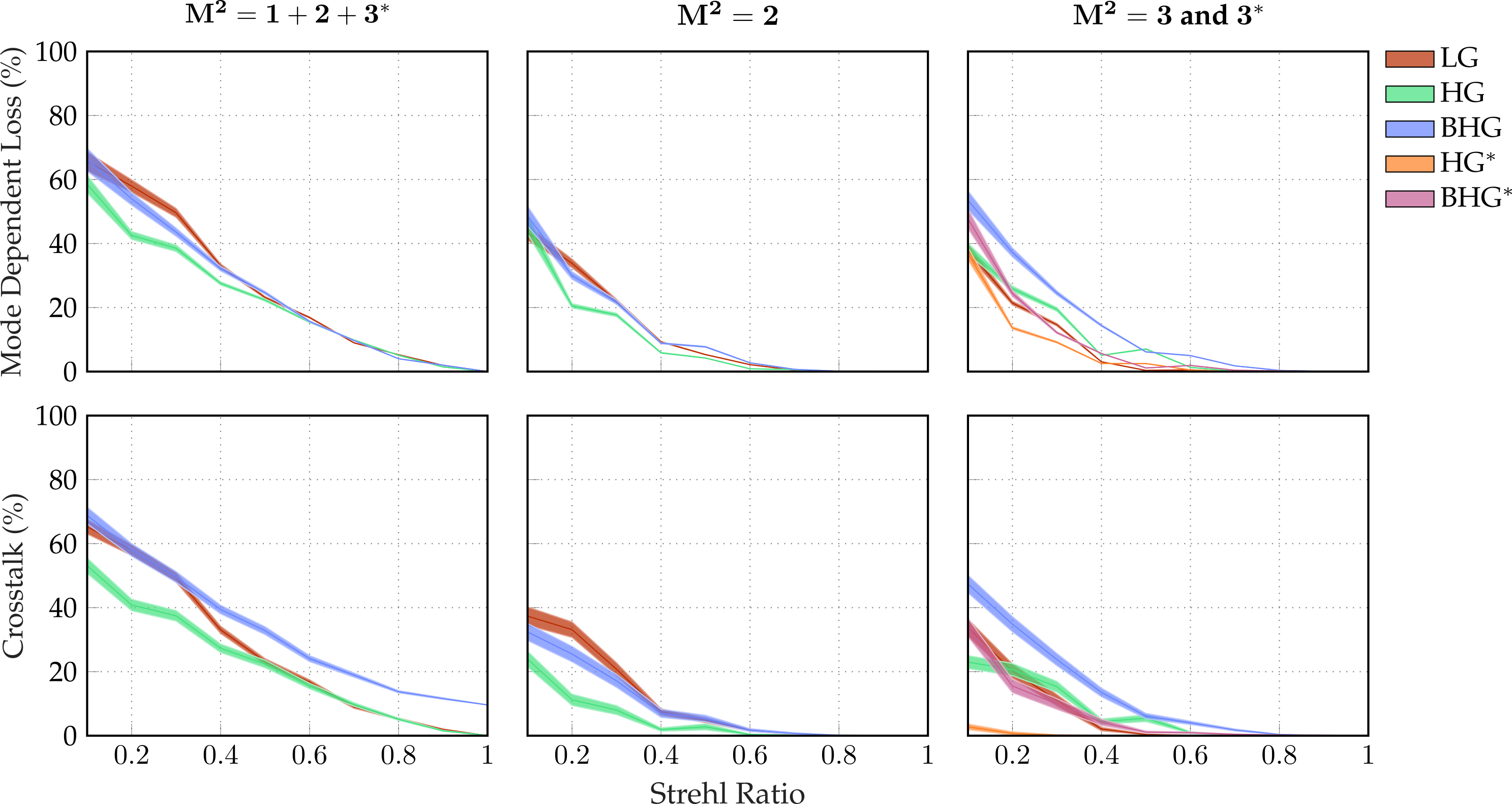}
  \caption{\label{fig:mdlcrosstalk}Mode Dependent Loss (top) and Crosstalk (bottom) of the LG, HG and Binary HG mode sets for different turbulence strengths form strong turbulence (SR=0.1) to no turbulence (SR=1.0). The shaded area of each curve represents the measurement error. The $3^*$ case excludes the HG$^1_1$ mode.}
\end{figure*}

\begin{figure}[th]
  \centering
      \includegraphics[width=.75\linewidth]{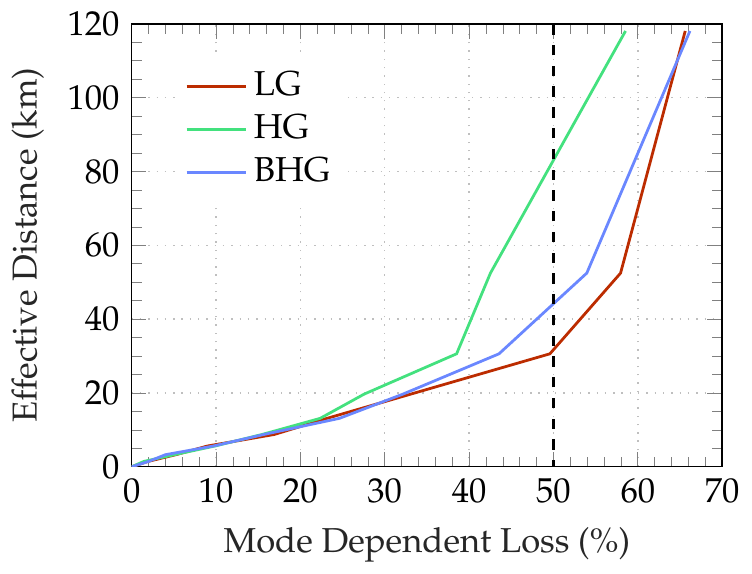}
  \caption{\label{fig:distance} Effective propagation distance ignoring divergence and atmospheric attenuation calculated from the Strehl Ratio using Eqs.~\ref{eq:sr} and \ref{eq:propagationFromSR} for the average ($M^2=1+2+3^*$) of each mode set. At the vertical dotted line the difference between HG and LG is 51~km and BHG and LG is 12~km. }
\end{figure}

The MDL results are shown at the top of Fig.~\ref{fig:mdlcrosstalk}. We show three cases, defined by the beam propagation factor, $M^2$. The Gaussian case, where $M^2=1$ is not shown explicitly as it corresponds to the measurement of the Strehl Ratio but it is included in the average case. We see that compared to LG modes, HG modes almost always exhibit a lower (better) or similar MDL. The BHG modes exhibit similar MDL to the LG modes, and are thus worse than HG modes, but this is expected because of their weaker orthogonality.

Interestingly, when the (B)HG$^1_1$ modes are included ($M^2=3$), the set exhibits significantly poorer performance in turbulence than without the (B)HG$^1_1$ mode ($M^2=3^*$). As visible in Fig.~\ref{fig:modes}, the (B)HG$^1_1$ is not tip or tilt invariant, making it more similar to an LG mode.

The crosstalk results shown at the bottom of Fig.~\ref{fig:mdlcrosstalk} agree with the MDL results in that HG modes exhibit lower crosstalk than LG modes. As expected, the BHG modes consistently show more crosstalk than either the HG or LG modes. Again, when the (B)HG$^1_1$ mode is excluded, the results agree with our hypothesis that HG modes are more resilient to turbulence than LG modes. 

Strangely, the percentage crosstalk of the HG modes excluding the HG$^1_1$ is very low. This result agrees with a visual inspection of Fig.~\ref{fig:verification}, where it is clear that in general the $M^2=3^*$ case experiences low crosstalk within the set, lending confidence to the result. While it has not been explicitly shown elsewhere, it is logical that the tip/tilt resilience of HG modes should extend to higher order modes with the same symmetry \cite{ndagano2017}. This resilience is visible in Fig.~\ref{fig:verification} in the crosstalk between the $M^2=2$ and $3^*$ sets.

The consequences of these results are important for FSO communications. Carefully chosen modes from the HG basis should result in superior MDM performance over modes from the LG basis, assuming the optical components such as the transmit and receive apertures have suitable geometry or are larger than the beams. The very low crosstalk results for HG modes with orthogonal symmetry is of particular interest. More important is the impact on FSO link distances. We see from Eqs.~\ref{eq:r0fromCn2} and \ref{eq:sr} that a lower Strehl Ratio is equivalent to a longer propagation distance, obviously ignoring atmospheric attenuation and beam divergence. Using MDL as the factor of comparison, we can find the effective propagation difference for each turbulence strength,
\begin{equation}
\label{eq:propagationFromSR}
L \approx \frac{0.060 \lambda^2}{C_n^2 r_0^{5/3}},
\end{equation}
and arbitrarily assuming $C_n^2=10^{-14}$~m$^{-2/3}$, which is a typical value \cite{Andrews2005}. The effective propagation distance for each mode set is plotted in Fig.~\ref{fig:distance}, where is it clear that HG and often BHG modes are superior to LG modes in terms of effective propagation distance for most MDL values. For instance, arbitrarily choosing a MDL of 50\% in a turbulence strength $C_n^2=10^{-14}$~m$^{-2/3}$, an LG mode could propagate 31~km whereas an HG mode would propagate 83~km. This is a significant increase in range. 

Practically, divergence and attenuation must be considered in a real-world FSO link. We have compared modes with the same beam propagation factor and so the divergence of the mode sets is comparable. Thus, while the realisable propagation distance would indeed be limited by practical factors, because HG modes experience a lower MDL than LG modes, they should enable longer range FSO communications. 

\section{Conclusion}
\label{sec:conc}
Atmospheric turbulence predominantly manifests as wandering of a FSO laser beam, resulting in fading (or MDL) as well as mode-crosstalk. Recent findings have shown that HG modes are resistant to tip and tilt aberrations \cite{Xie2015,Pang2018HGLG,ndagano2017}. Given the complex phase structure of LG and HG modes, it was unknown whether the higher order effects of turbulence would overcome the tip/tilt resilience of HG modes over LG modes. In this work we experimentally show that HG and binary HG modes are significantly more resilient to Kolmogorov turbulence than LG beams with the same beam propagation factor ($M^2$). Lower crosstalk results in higher capacity mode division multiplexed systems, but more importantly, the lower MDL experienced by HG modes over LG modes means that they can propagate significantly further. At a MDL of 50\%, we show a 167\% increase in theoretical range of non-symmetric HG modes over similar order LG modes. FSO communications are a possible technology which can be used to bridge the digital divide in Africa where large geographical distances must be traversed, provided the range of such systems makes this course economical over the cost of laying fibre. This work is a significant step towards improving the range of FSO links.



\ifCLASSOPTIONcaptionsoff
  \newpage
\fi



%


\end{document}